\documentclass[conference]{IEEEtran}
\IEEEoverridecommandlockouts
\usepackage{cite}
\usepackage{amsmath,amssymb,amsfonts}
\usepackage{algorithmic}
\usepackage{graphicx}
\usepackage{textcomp}
\usepackage{xcolor}
\usepackage{multirow}
\usepackage{verbatim}
\usepackage{float}
\usepackage{cancel}
\usepackage{tikz}
\usepackage[hidelinks]{hyperref}
\usetikzlibrary{shapes.geometric, arrows}
\newcommand{\vectext}{\text{vec}}

\def\BibTeX{{\rm B\kern-.05em{\sc i\kern-.025em b}\kern-.08em
    T\kern-.1667em\lower.7ex\hbox{E}\kern-.125emX}}
\begin{document} 

% \title{Comparative Analysis of EIV Solution Techniques in Presence of Non-Gaussian Noises: A Case-Study of Time-Synchronized Line Parameter Estimation\\
% }

% \author{\IEEEauthorblockN{Anushka Sharma}
% \IEEEauthorblockA{Electrical and Electronics Engineering\\
% \textit{Birla Institute of Technology}\\
% Patna, India \\
% nushka0912@gmail.com}
% \and
% \IEEEauthorblockN{Antos Cheeramban Varghese}
% Student Member, IEEE \\
% \IEEEauthorblockA{School of ECEE\\
% \textit{Arizona State University}\\
% Tempe, AZ \\
% avarghe6@asu.edu}
% \and
% \IEEEauthorblockN{Anamitra Pal}
% Senior Member, IEEE
% \IEEEauthorblockA{School of ECEE\\
% \textit{Arizona State University}\\
% Tempe, AZ \\
% anamitra.pal@asu.edu}
% }

% \title{Time-Synchronized Estimation of Line Parameters in Presence of Non-Gaussian Measurement Noise}
%\title{Phasor Measurement Unit-based Line Parameter Estimation in Presence of Non-Gaussian Noise}
\title{\textcolor{black}{Comparative Analysis of Information Theoretic and Statistical Methods for Line Parameter Estimation}}

\author{\IEEEauthorblockN{Anushka Sharma, \textit{Undergraduate Student Member,}\\
 \textit{IEEE}}
\IEEEauthorblockA{Electrical and Electronics Engineering\\
\textit{Birla Institute of Technology}, Patna, India \\
btech15200.20@bitmesra.ac.in}
\and
\IEEEauthorblockN{Antos C. Varghese, \textit{Graduate Student Member,}\\
 \textit{IEEE}, and Anamitra Pal, \textit{Senior Member, IEEE}}
\IEEEauthorblockA{School of Electrical, Computer, and Energy Engineering\\
\textit{Arizona State University}, Tempe, AZ \\
avarghe6@asu.edu, anamitra.pal@asu.edu}
\and
% \IEEEauthorblockN{Anamitra Pal}
% Senior Member, IEEE
% \IEEEauthorblockA{School of ECEE\\
% \textit{Arizona State University}\\
% Tempe, AZ \\
% anamitra.pal@asu.edu}
\thanks{This work was supported in part by the National Science Foundation
(NSF) Grant ECCS 2145063.}
}

\maketitle

\begin{abstract}
Recent studies indicate that the noise characteristics of phasor measurement units (PMUs) \textcolor{black}{can be} more accurately described by non-Gaussian distributions.
% , rather than the traditionally assumed Gaussian. 
Consequently, estimation techniques based on Gaussian noise assumptions may produce poor results with PMU data.
% Estimation techniques rooted in Gaussian noise assumptions tend to produce imprecise results under these conditions. 
This paper considers the PMU-based line parameter estimation (LPE) problem, and investigates the performance of four state-of-the-art techniques in solving this problem in presence of non-Gaussian measurement noise.
% This paper explains four advanced methods that are tailor-made for non-Gaussian measurement noises: Minimum Total Error Entropy (MTEE), Maximum Total Correntropy (MTC), Constrained MTC (CMTC), and Errors-in-variables for Gaussian mixture model-Lagrangian-Expectation maximization (EGLE). In-depth performance comparisons of these methods in various scenarios are carried out using the line parameter Estimation problem as the case study. 
% The results 
The rigorous comparative analysis highlights the merits and demerits of each technique w.r.t. the LPE problem, and identifies conditions under which they are expected to give good results.
% An in-depth comparison of the techniques
% indicates that the techniques are suitable under minimal tuning, knowledge of bounded variation, and higher accuracy scenarios. 
\end{abstract}

\begin{IEEEkeywords}
Line parameter estimation (LPE), Non-Gaussian measurement noise, Phasor measurement unit (PMU) 
\end{IEEEkeywords}

\section{Introduction}
\label{Intro}
%In the dynamic landscape of contemporary power systems, precise line parameter estimation serves as the linchpin for operational reliability and efficiency. However, the intricate complexities within power networks, alongside external influences, introduce uncertainties in data collected from synchronized measurement devices. To confront this challenge, Errors-in-Variables (EIV) models have emerged as adept tools, capable of providing accurate parameter estimates by accounting for the inherent imprecision present in both input and output data.

In the dynamic landscape of modern
% contemporary 
power systems, accurate line parameters
% precise line parameter estimation 
serve as the linchpin for operational reliability and efficiency. 
Hence, the ability to estimate them whenever needed using phasor measurement units (PMUs) placed on both ends of the lines was a significant breakthrough for synchrophasor technology \cite{mansani2018estimation,mishra2015kalman}.
PMUs provide voltage and current phasor measurements, from which line parameters can be found using Ohm's law and Kirchhoff's laws. 
However, since both voltage and current measurements from PMUs are noisy, the conventional least squares (LS) approach, which only assumes noise in the dependent variable, is unable to give accurate parameter estimates.
%However, since both voltage and current measurements from PMUs are noisy, due to potential errors arising due to different system operating conditions, aging process of instrument transformers, incorrect time synchronization, errors introduced by the phasor estimation algorithm, varying communication channel noises, and/or cyber-attacks such as eavesdropping, global positioning, system (GPS) spoofing and data tampering. The conventional least squares (LS) approach, which only assumes noise in the dependent variable, is unable to give accurate parameter estimates.
% \textcolor{black}{However, since voltage and current measurements obtained from PMUs are subject to noise stemming from various factors including different system operating conditions, aging of instrument transformers, incorrect time synchronization, errors introduced by the phasor estimation algorithm, fluctuations in communication channel noise, and potential cyber-attacks such as eavesdropping, global positioning system (GPS) spoofing, and data tampering. The conventional least squares (LS) approach, which only assumes noise in the dependent variable, is unable to give accurate parameter estimates.}
% However, the intricate complexities within power networks, alongside external influences, introduce uncertainties in data collected from synchronized measurement devices. 
To confront this challenge, errors-in-variables (EIV) models have emerged as adept tools, capable of providing accurate parameter estimates by accounting for the imprecision present in both dependent and independent variables.
The line parameter estimation (LPE) problem has been solved as an EIV problem in \cite{albuquerque2021estimation,wehenkel2020parameter,dasgupta2013line,ding2011transmission}.
% \cite{lin2019synchrophasor, wehenkel2020parameter, xue2019linear, gupta2021compound}.

% The traditional approach for solving EIV problems is by using the total least squares (TLS) method, and it has already been used to solve the line parameter estimation (LPE) problem in \cite{ding2011transmission}. 

Traditionally, EIV problems are solved using the total least squares (TLS) method.
Similar to LS, TLS is also designed to give optimal estimates when the noises in the dependent and independent variables follow a Gaussian distribution. 
This is because both LS and TLS rely on second-order statistics and aim to minimize the \textit{sum of squares error} (SSE). 
However, it has recently been discovered through rigorous statistical analysis that the noise in the PMU measurements \textcolor{black}{can have} a non-Gaussian distribution \cite{salls2021statistical},\textcolor{black}{\cite{castello2022statistical}}. 
\textcolor{black}{The non-Gaussianity can stem from diverse operating conditions, aging process of instrument transformers, incorrect time synchronization, errors introduced by phasor estimation algorithms, and/or varying communication channel noises \cite{ahmad2019statistical}.}
Consequently, there is a pressing need to delve into advanced estimation techniques for solving EIV problems that are capable of handling non-Gaussian noise in the dependent and independent variables.
% A pivotal hurdle lies in the presence of non-Gaussian noise, a factor that significantly complicates parameter estimation. Traditional methods, grounded in the assumption of Gaussian noise characteristics, often falter when confronted with the intricacies posed by non-Gaussian noise sources. Consequently, there exists a pressing need to delve into advanced EIV solution techniques, engineered to robustly tackle the intricate distribution patterns intrinsic to non-Gaussian noise scenarios.

% Recent advancements in the signal processing domain
% have showcased the ascendancy of Information Theoretic Learning (ITL)--based algorithms in tackling non-Gaussian measurement noise from an estimation perspective.
% %, fortified by the capabilities of Phasor Measurement Units (PMUs). 
% % These algorithms have revolutionized real-time impedance parameter estimation. 
% Among them, notable contenders include the 
% % Least Square approach (LS), Total Least Square (TLS),
% minimum total error entropy (MTEE) and maximum total correntropy (MTC). 
% % Both of these algorithms have significantly elevated the precision of parameter estimation methodologies.

This paper starts by analyzing four approaches that have been proposed recently to perform linear estimation\footnote{Note that LPE when done using PMU measurements can be expressed as a linear estimation problem.} 
for EIV problems in presence of non-Gaussian noise. 
% in the dependent and independent variables.
Minimum total error entropy (MTEE) \cite{shen2015minimum} and maximum total correntropy (MTC) \cite{wang2017maximum} are based on information theoretic learning (ITL) and minimize the ``net error" without any additional constraints.
Constrained MTC (CMTC), which is also based on ITL, solves the MTC problem with equality constraints \cite{QIAN2021107903}.
% The first two methods, namely, minimum total error entropy (MTEE) \cite{shen2015minimum} and maximum total correntropy (MTC) \cite{wang2017maximum}, are based on information theoretic learning (ITL), and minimize the ``net error" without any additional constraints.
% The third technique is the constrained MTC (CMTC), which is also based on ITL, and solves the MTC problem with equality constraints \cite{QIAN2021107903}.
The fourth approach, termed EGLE (EIV for Gaussian mixture model-Lagrange multipliers-Expectation maximization) \cite{varghese2022transmission}, uses robust statistics to perform noise and parameter estimation simultaneously.
Subsequently, the paper applies these state-of-the-art techniques to solve the PMU-based LPE problem under diverse noise conditions.
% The objective of this paper is to apply these state-of-the-art estimation techniques, which are specifically designed for non-Gaussian noise environments, to the time-synchronized LPE problem.
% Furthermore, we conduct a rigorous comparative analysis to highlight the merits and demerits of each technique w.r.t. this problem. 
The IEEE 118-bus system is used as the test system for this analysis.  
% The objective is to conduct rigorous comparison studies and highlight the merits and demerits of each method.  
% The rigorous comparative analysis highlights the merits and demerits of each technique w.r.t. this problem. 
The results indicate that (a) with appropriate hyperparameter tuning, MTC/CMTC can give good results; (b) along with extensive hyperparameter tuning, MTEE also needs sufficient time/compute resources to give acceptable results; (c) EGLE gives fast and reasonably accurate line parameter estimates with minimal tuning. 

The rest of the paper is structured as follows: Section \ref{sec:SOTA} provides a brief overview of the MTEE, MTC, CMTC, and EGLE algorithms. Section \ref{sec:lpe} explains the PMU-based LPE problem. The performances of the four techniques for the PMU-based LPE problem are described in Section \ref{sec:perf}. Finally, the conclusion is provided in Section \ref{sec:conc}. 

% \subsection{Notation}
% \noindent \textit{Notation:} 
\textit{Notation:} 
% In this paper, uppercase and lowercase variables denote matrix and vectors, respectively. 
The superscript $\tilde{(.)}$ denotes noisy data, the subscript $(.)_{true}$ denotes the true value, and the superscript $\hat{(.)}$ denotes the estimated value. 
% $n$ and $p$ denote the number of samples (time instants) and the number of parameters to be estimated, respectively.

\section{Solving EIV Problems with Non-Gaussian Noise}
\label{sec:SOTA}

%\subsection{Noise Modeling}

A linear estimation problem can be written as

\begin{equation}
\label{True_linear_system}
y_{true} = X_{true} \: w_{true}
\end{equation}
% \eqref{True_linear_system} represents a linear model where the
% %current
% the true value of the dependent variable
% \begin{math} y_{true} \in \mathbb{R}^{n \times 1} \end{math} is expressed as a product of the 
% %voltage
% true value of the independent variable \begin{math} X_{true} \in \mathbb{R}^{n \times p} \end{math} and the parameter vector \begin{math} w_{true} \in \mathbb{R}^{p \times 1}\end{math}; $n$ and $p$ denoting the number of samples (time instants) and the number of parameters to be estimated, respectively. 
where, \begin{math} y_{true} \in \mathbb{R}^{n \times 1} \end{math} is the true value of the dependent variable, \begin{math} X_{true} \in \mathbb{R}^{n \times p} \end{math} is the true value of the independent variable, \begin{math} w_{true} \in \mathbb{R}^{p \times 1}\end{math} is the true value of the parameter to be estimated, and $n$ and $p$ denote the number of samples and the number of parameters to be estimated, respectively.
% In real-life situations, measurement noise is inevitable, and we often have access to noisy measurements. In this scenario, the linear system with noisy measurement can only be represented as
In presence of measurement noise, \eqref{True_linear_system} becomes
\begin{equation}
\label{Noisy_linear_system}
\tilde{y}  \approx \tilde{X}  \: w
\end{equation}
where, $\tilde{y}$ and $\tilde{X}$ represent the noisy version of dependent and independent variables, respectively, and $w$ is the parameter to be estimated.
The goal is to solve the linear estimation problem by finding the $\hat{w}$ that best fits the observed data.
% The goal is to estimate the parameters of the linear system \begin{math} w \end{math} based on observed (noisy) values of the dependent and independent variables.
% In other words, the objective is to find the optimal values of \begin{math} w \end{math} that best fit the observed data. 
Several techniques, such as 
% LS, 
TLS, MTEE, MTC, CMTC, and EGLE, 
can achieve this goal. However, the estimation accuracy varies depending on the noise characteristics and the assumptions on which the techniques are built.

% LS 
% % and TLS are already well-known in the power systems community. The former 
% gives the optimal estimate when Gaussian noise is only present in the dependent variable and the independent variable is noise-free (i.e., $\tilde{X}=X_{true}$). The optimal estimate obtained using LS ($\hat{w}_{LS}$) is shown in \eqref{w_LS}.

% \begin{equation}
% \label{w_LS}
% \begin{aligned}
% \hat{w}_{LS} &= (\tilde{X}^T \tilde{X})^{-1} (\tilde{X}^T \tilde{y}).  \\
% \end{aligned}  
% \end{equation}

TLS assumes Gaussian noise in both dependent and independent variables, and computes the optimal parameter estimate by solving \eqref{TLS_eq} \cite{markovsky2007overview},
% the following equation,
\begin{equation}
\label{TLS_eq}
    \hat{w}_{TLS} = [d_{qq}]^{-1} [d_{pq}]
\end{equation}
where, $d_{pq}$ is the vector of first $p$ elements and $d_{qq}$ is the $(p+1)^{th}$ element, respectively, of the $(p+1)^{th}$ column of the matrix of right singular vectors of the singular value decomposition of $[\tilde{X}~\tilde{y}]$.
However, the performance of TLS deteriorates
% The performance of both LS and TLS deteriorate 
in presence of non-Gaussian noise in the measurements that comprise the dependent and independent variables. 
Four techniques that are, by definition, immune to non-Gaussian noises in the EIV context, are described next.

\subsection{Minimum Total Error Entropy (MTEE)} 
MTEE involves minimizing the quadratic Renyi's entropy of the total error, $e^{tot}$, to obtain accurate parameter estimates, where $e^{tot}$ is mathematically expressed as
\begin{equation}e^{tot}= \frac{\tilde{y}-\tilde{X}w}{(\sqrt{\lVert w^2 \rVert+\epsilon_0^{-2})}}.
\label{e_tot}
\end{equation}

In \eqref{e_tot}, $\epsilon_0$ is the square-root of the ratio of the noise intensity of the independent variable to the noise intensity of the dependent variable. If the exact values of noise intensity are not known, but it can be assumed that the noise intensity of the independent variable is comparable to that of the dependent variable\footnote{This can happen when noises in both the variables are coming from sensors that satisfy the same standards; e.g., IEC/IEEE 60255-118-1:2018 Standard for PMU measurements \cite{IEEE_IEC2018PMU_Std}.}, then we can set $\epsilon_0 = 1$. 
The estimator for Renyi's entropy, denoted by $\hat{V}_2(e^{tot})$, 
% which also corresponds to the quadratic information potential, 
can now be expressed as
\begin{equation}
J_{\mathrm{MTEE}} = \hat{V}_2(e^{tot}) = \frac{1}{n^2}\sum_{i=1}^{n}\sum_{j=1}^{n}G_{\sigma\sqrt{2}}(e_j^{tot}-e_i^{tot})
\label{MTEE_estimator}
\end{equation}
where, $G_{\sigma}$ represents a Gaussian kernel function with a kernel width of $\sigma$. % and \begin{math} e^{tot} \end{math} can be computed by using the following expression,\\
% (let \begin{math}\epsilon_0 = \frac{\sigma_{in}}{\sigma_{out}} \end{math})
% \begin{equation}e^{tot}= \frac{\tilde{Y}-\tilde{X}w}{(\sqrt{\lVert w^2 \rVert+\epsilon_0^{-2})}}\end{equation}
The MTEE approach aims to iteratively adjust $w$ to maximize $\hat{V}_2(e^{tot})$, and obtain accurate parameter estimates in the process.
The necessary condition to find the maximum value of the estimator shown in \eqref{MTEE_estimator} is given in \eqref{MTEE_update}, where $\Delta e_{ij}^{tot} = e_{i}^{tot} - e_{j}^{tot}$, $\Delta \tilde{x}_{ij} = \tilde{x}_{i} - \tilde{x}_{j}$, where $\tilde{x}_{i}$ is the $i^{th}$ row of $\tilde{X}$, $\bar\epsilon = \epsilon_0^{-2}$, and the superscript $tot$ is suppressed to avoid notational clutter.
\begin{equation}
\begin{aligned}
g_{\mathrm{MTEE}} = \frac{\partial \hat{V}_2(e)}{\partial w} &= \frac{1}{\sigma^2n^2}\cdot \sum_{i=1}^{n}\sum_{j=1}^{n}G_{\sigma}(\Delta e_{ij}) \\ & \times \left(\frac{\Delta e_{ij}^2w}{w^Tw + \bar\epsilon}  + \frac{\Delta e_{ij}\Delta \tilde{x}_{ij}^T}{\sqrt{w^Tw + \bar\epsilon}} \right).
\label{MTEE_update}
\end{aligned}
\end{equation}

% where, $\bar\epsilon$ is given by $\epsilon_0^{-2}$.
Using \eqref{MTEE_update}, the parameter update step can be written in the form of the steepest ascent algorithm as shown below \cite{shen2015minimum},
% in the steepest ascent algorithm for MTEE solution methodology is described as

\begin{equation}
w_{r+1} = w_r + \mu \times g_{\mathrm{MTEE}}\bigg|_{w_r}
\label{steepest ascent}
\end{equation}
where, $r$ and $\mu$ denote the iteration number and the learning step size of the steepest ascent algorithm, respectively.

% This approach aims to iteratively adjust the parameter vector $w$ to maximize Renyi's entropy estimator $\hat{V}_2(e)$ and obtain accurate parameter estimates in the process.

\subsection{Maximum Total Correntropy (MTC)}
% The parameter estimation in the Maximum total correntropy method is done by maximizing the following correntropy-based cost function defined as
In the MTC technique, the following correntropy-based cost function is maximized:
\begin{equation} J_{\mathrm{MTC}} = \mathbb{E} \left[\exp \left(-\frac{e_{i}^2}{2\sigma_{\mathrm{MTC}}^2 \lVert \bar w \rVert^2}  \right)\right].
\label{MTC_cost}
\end{equation}
% \begin{equation} J_{MTC} = E\left[exp\left(-\frac{e^2(i)}{2\sigma_{MTC}^2 \lVert \bar w \rVert^2}  \right)\right]
% \label{MTC_cost}
% \end{equation}

In \eqref{MTC_cost}, $e_{i}=\tilde{y}_i-\tilde{x}_iw$, $\sigma_{\mathrm{MTC}}$ is the Gaussian kernel width, and 
$\bar w = [\epsilon_{0}^{-1}~-\!w]$.
% is the modified augmented parameter vector
% Here, the modified augmented weight vector, represented as $\bar w$, is determined by taking the transpose of the expression obtained by subtracting the transposed weight vector $w^T$ from the square root of the ratio of the output variance ($\sigma_o^2$) to the input variance ($\sigma_i^2$). The computed deviation at iteration $i$, denoted as $e(i)$, is given by $\tilde{y}-w^T\tilde{X}$. Furthermore, $\sigma_{MTC}$ represents kernel width.
The objective of the MTC technique is to minimize the mean of non-linear weighted squared residuals.
% For the gradient descent method, the partial derivative of the cost function is defined as
This can be done using the gradient descent method by computing the partial derivative of the cost function shown in \eqref{MTC_cost}, as written below
% \begin{equation}
% \begin{split}
% g_{\mathrm{MTC}} = \frac{\partial {J_{\mathrm{MTC}}(w)}}{\partial{w}} &= \frac{1}{n\sigma_{\mathrm{MTC}}^2} \sum_{i=1}^{n} \left[\exp\left(\frac{-e^2_{i}}{2\sigma^2_{\mathrm{MTC}}\lVert \bar w \rVert^2} \right) \right. \\
% &\quad \left. \times \frac{\left(\lVert \bar w \rVert^2 e_i\tilde{x}_i^T + e^2_{i} w\right)}{\lVert \bar w \rVert^4} \right]
% \end{split}
% \label{gradient_mtc}
% \end{equation}
\vspace{-3mm}

\begin{equation}
\begin{split}
&g_{\mathrm{MTC}} = \frac{\partial {J_{\mathrm{MTC}}(w)}}{\partial{w}} = \frac{1}{n\sigma_{\mathrm{MTC}}^2} \times
\\
&\quad
\sum_{i=1}^{n} \left[\exp\left(\frac{-e^2_{i}}{2\sigma^2_{\mathrm{MTC}}\lVert \bar w \rVert^2} \right) \right.  \left. \times \frac{\left(\lVert \bar w \rVert^2 e_i\tilde{x}_i^T + e^2_{i} w\right)}{\lVert \bar w \rVert^4} \right].
\end{split}
\label{gradient_mtc}
\end{equation}

At iteration $r$, the parameter update of the gradient-based MTC algorithm is obtained using \eqref{gradient_mtc}, as shown below \cite{wang2017maximum}:
\begin{equation} w_{r+1} = w_r + \eta \times g_{\mathrm{MTC}}|_{w_r}
\end{equation}
where, $\eta = \mu \cdot \sigma_{\mathrm{MTC}}^2$.
% , and is called the learning rate.

\subsection{Constrained Maximum Total Correntropy (CMTC)}
\label{CMTC_subsection}
%The MTC algorithm can be modified to incorporate the constraint $Y_1 + Y_3 = 0$  and a Constrained MTC algorithm can be introduced.
% The line parameter estimation modeling explained in section \ref{sec:lpe} has a linear equality constraint. An advanced method that could incorporate equality constraint to the naive MTC could leverage this additional information for improved estimation. The constrained MTC does exactly this.   
The CMTC technique solves the MTC problem in presence of equality constraints.
The mathematical formulation of this linear estimation problem is given below,
%In the existing EIV equation.
%\begin{equation} y = X w
%\end{equation}
%The constraint can be incorporated by using the following expression
%\begin{equation} C^T w = f
%\end{equation}
% The linear system along with an equality constraint for CMTC can be expressed as
\begin{equation}
\label{EIV_noisy_linear_system_indep_variable}
\begin{aligned}
%  (y_{true}+\tilde{y}) &= (X_{true}+\tilde{X})  w \\
% \tilde{y} &= y_{true} + y_{n} \\
% \tilde{X} &= X_{true} + {X}_n \\
\tilde{y}  \approx \tilde{X}  \: w \\
C^T w = f
\end{aligned}  
\end{equation}
in which the second sub-equation is the equality constraint.
Through the utilization of the Lagrange multiplier method, the cost function of CMTC is acquired as shown below,
\begin{equation} J_{\mathrm{CMTC}} = \mathbb{E} \left[G_\sigma\left(\frac{e_i}{\sqrt{\bar w^T\bar w}}  \right)\right] + \lambda^T(C^T w - f)
\end{equation}
where, $\lambda$ is the Lagrange multiplier, 
which for the $r^{th}$ iteration, is given by,
% and is given by
% \begin{equation}
% \lambda_r = \frac{1}{\eta} \cdot (C^TC)^{-1}(f - C^T w_{r-1} - C^T\eta g_{\mathrm{MTC}} (w_{r-1}))
% \end{equation}
\begin{equation}
\lambda_{r+1} = \frac{1}{\eta} \cdot (C^TC)^{-1}(f - C^T w_{r} - C^T\eta g_{\mathrm{MTC}} (w_{r})).
\end{equation}

By employing the stochastic gradient descent, we derive the updated parameter estimate of CMTC as shown below \cite{QIAN2021107903}:
\begin{equation} w_{r+1} = w_r + \eta \times g_{\mathrm{MTC}}|_{w_r} + \eta C\lambda_{r+1}.
\end{equation}

% As the equality constraint is also incorporated along with traditional MTC capabilities, CMTC is expected to perform better than MTC for constrained optimization problems albeit at higher computation cost. 

\subsection{EIV for Gaussian mixture model-Lagrange multipliers-Expectation maximization (EGLE)}

The preceding subsections described three
% elaborated on 
estimation methods that were based on ITL. In contrast, the EGLE algorithm harnesses \textit{robust statistics} to address potential non-Gaussian noise in EIV problems.
Another key distinction is that MTEE, MTC, and CMTC were noise-model agnostic, while EGLE explicitly expresses the noise in terms of Gaussian mixture models (GMMs).
 %Let the noise in the dependent and independent variable be defined as shown below
 \begin{comment}
EGLE is designed for a generic noise environment defined as
\begin{equation}
\label{EIV_noisy}
\begin{aligned}
\tilde{y} &= y_{true} + y_{n} \\
\tilde{X} &= X_{true} + {X}_n \\
\end{aligned}  
\end{equation}
where, $y_{n}$ and $X_{n}$ can have non-Gaussian characteristics.
\end{comment}
The mathematical basis for the working of EGLE is that, in the presence of noises modeled as GMMs, minimizing the sum of squares of \textit{standardized} error (SSSE) with respect to the associated Gaussian component in GMM yields optimal parameter estimates (rather than minimizing SSE as done in TLS).
The $g^{th}$ Gaussian component of the \textit{standardized} error in the dependent and independent variable (denoted by $e_{g_S}$ in the subscript) can be written in the form:
% using expectation maximization (EM) 
\begin{subequations}
\label{EGLE_noise}
    \begin{align}
    y_{e_{g_S}} &= y_{\Sigma_g}^{-\frac{1}{2}} \: (y_{e_{g}} - y_{\mu_g})
    \\
    X_{e_{g_S}} &= X_{\Sigma_g}^{-\frac{1}{2}} \: (X_{e_{g}} - X_{\mu_g}) 
    \end{align}
\end{subequations}
where, 
for the corresponding variable, 
the subscripts $e_g$ denote the measurement noise, and 
$\mu_g$ and $\Sigma_g$ denote the mean and covariance, of the $g^{th}$ Gaussian component,
% of the corresponding dependent/independent variable, 
respectively. 
% After finding the noise expressions, expectation maximization is done to determine the new GMM parameters 
% % (means, standard deviations, and weights)
% % parameters: $\mu_g$, $\sigma_g$, and weights ($\Lambda_g$), 
% that can approximate $y_{e_{g_S}}$ and $X_{e_{g_S}}$.
% % the noises.
%These GMM parameters act as inputs to the subsequent parameter estimation step in addition to the noisy measurements. 
%  \begin{equation}
%      y_{n_{g_S}} = [\Sigma_{y_g}^{-\frac{1}{2}} \: (\tilde{y}_{n_{g}} - \mu_{y_g})]
%  \end{equation}
% % Similarly, the $g^{th}$ Gaussian component of the standardized independent variable noise vector can be denoted as
%  \begin{equation}
%     X_{n_{g_S}} = [\Sigma_{X_g}^{-\frac{1}{2}} \: (\tilde{X}_{n_{g}} - \mu_{X_g})]
%  \end{equation}
% Note that $y_{e_{g_N}}^T  y_{e_{g_N}}$ denotes the SSSE in $y_{e_{g}}$. 
% Being a matrix, $X_{n_{g_S}}$, is first converted to a column vector by vectorization operation.  
The overall minimization problem for optimal noise and parameter estimation by EGLE is mathematically expressed as
%However, as $X_{n_{g_S}}$ is a matrix, it has to be first converted to a column vector form before it can be converted to an SSSE form. Let $\vectext(X_{n_{g_S}})$ denote the vectorization operation of the matrix $X_{e_{g_N}}$, where the matrix is converted to a column vector by stacking the columns of the matrix on top of each other.
%Then, using the properties of the Kronecker product, $\otimes$, the 
%resulting minimization problem for optimal parameter estimation for the EIV problem can be mathematically described by:
\begin{equation}
\label{OpParaEIV}
\begin{aligned}
    J_{\mathrm{EGLE}} &= \arg \min_{\substack{w }} \:\:  \sum_{g=1}^{m} \frac{1}{2} ( y_{e_{g_S}})^T ( y_{e_{g_S}}) \\
      &+ \sum_{g=1}^{m} \frac{1}{2} (\vectext( X_{e_{g_S}}))^T  (\vectext( X_{e_{g_S}}))\\
    \text{s.t. } & [\tilde{y}_{g}  - y_{e_{g}} ] =  (w^T \otimes I_{n_g}) ( \vectext(\tilde{X}_{g})- \vectext(X_{e_{g}}) )\\
    &\:\: \forall g \in [1, \dots, m].
%    \text{s.t.}\:\: (c-c_e) &= (D - D_e) x
\end{aligned}
\end{equation}
where, $\mathrm{vec}$ denotes the vectorization operation, $\otimes$ is the Kronecker product, $I_{n_g}$ is the identity matrix of size $n_g$, where $n_g$ is the number of samples belonging to the $g^{th}$ Gaussian component, and $m$ is the number of Gaussian components.
\begin{comment}
Using the method of Lagrange multipliers, \eqref{OpParaEIV} can be written as an unconstrained objective function as

\begin{equation}
\begin{aligned}
    \phi &= \arg \min_{\substack{x }} \:\:  \sum_{g=1}^{m} \frac{1}{2} ( c_{e_{g_N}})^T ( c_{e_{g_N}}) \\
      &+ \sum_{j=1}^{m} \frac{1}{2} (\vectext( D_{e_{g_N}}))^T  (\vectext( D_{e_{g_N}}))\\
  &+ \sum_{g=1}^{m} [c_{g} - (x^T \otimes I_{n_g}) ( \vectext(D_{g}) - \vectext(D_{e_{g}})) - c_{e_{g}} ]^T \lambda_g\\
%    \text{s.t.}\:\: (c-c_e) &= (D - D_e) x
\end{aligned}
\end{equation}
\end{comment}
Expressing \eqref{OpParaEIV} as an unconstrained optimization problem using Lagrange multipliers ($\alpha_g$), and substituting $y_{e_{g_S}}$ and $X_{e_{g_S}}$ from \eqref{EGLE_noise}, the optimal parameter estimate is obtained by solving
% for this objective function is the solution to 
the following system of non-linear equations:
\begin{equation}
\label{Gen_EIV_Solution1}
    \begin{aligned}
      f(w) = \sum_{g=1}^{m} (\tilde{X}_{g} - X_{e_{g}})^T \alpha_g  &= 0 \\
    \end{aligned}
\end{equation}
where,
\vspace{-3mm}
% \begin{equation}
% \label{Gen_EIV_Solution_sub_variables}
%     \begin{aligned}
%         \lambda_g &= (\Sigma_{ net g})^{-1} ( c_{g} - D_{g} x  - \mu_{ net g}). \\
%         \mu _{net g} &= \mu_{c_g} - \sum\limits_{j=1}^p \mu_{D_g} \times {x}_j \\
%         \Sigma_{ net g} &= \Sigma_{c_g} + \sum\limits_{j=1}^p \Sigma_{D_g} \times {x}_j^2.\\
%     \end{aligned}
% \end{equation}
\begin{subequations}
\label{Gen_EIV_Solution_sub_variables}
    \begin{align}
        \alpha_g  &= (\Gamma_{\Sigma_g})^{-1} ( \tilde{y}_{g} - \tilde{X}_{g} w  - \Gamma_{\mu_g} )\\
       \Gamma_{\mu_g} &= y_{\mu_g} - \sum\limits_{j=1}^p X_{\mu_g} {w}_j \\
        \Gamma_{\Sigma_g} &= y_{\Sigma_g} + \sum\limits_{j=1}^p X_{\Sigma_g} {w}_j^2.
    \end{align}
\end{subequations}

% In \eqref{Gen_EIV_Solution_sub_variables}, $j$ denotes the $j^{th}$ column for $X$ and $j^{th}$ parameter for $w$. 
% The optimal parameters is found by solving 
The above-mentioned system of non-linear equations 
can be solved using the standard Newton's method as shown below:
% (or Newton's method with equality constraints, when such constraints are present):
% \begin{equation}
%        \label{NM_Updation_1st_deri}
%            \begin{aligned}    
%           \begin{bmatrix}
%              w^{(r+1)}      
%           \end{bmatrix} = \begin{bmatrix}
%              w^{(r)}      
%           \end{bmatrix} - \begin{bmatrix}
%             \text{Jac}(f(w^{(r)}))      
%           \end{bmatrix}^{-1}   \begin{bmatrix}
%              f(w^{(r)})     
%           \end{bmatrix}
%         \end{aligned}
% \end{equation} 
\begin{equation}
       \label{NM_Updation_1st_deri}
           \begin{aligned}  
             w^{(r+1)}      
           = 
             w^{(r)}      
           - \begin{bmatrix}
            \text{Jac}(f(w^{(r)}))      
          \end{bmatrix}^{-1}   
             f(w^{(r)})   
        \end{aligned}
\end{equation} 
where, $\text{Jac}(f(w^{(r)}))$ is the Jacobian of $f(w^{(r)})$ at $r^{th}$ iteration.
When equality constraints are present, the system of non-linear equations will be solved using the Newton's method with equality constraints.
The components of the noise estimates of $y_e$ and $X_e$ are obtained by solving \eqref{EIV-noise-estimates}, where ${x_e}_{g_{j}}$ and ${x_{\mu}}_{g_{j}}$ denote the $j^{th}$ column of $X_{e_{g}}$ and $X_{\mu_g}$, respectively, and $w_j$ is the $j^{th}$ parameter of $w$.
\vspace{-1mm}
\begin{subequations}
\label{EIV-noise-estimates}
    \begin{align}
        \hat{y}_{e_{g}} &= y_{\Sigma_g} \alpha_g + y_{\mu_g}\\
      \hat{x_e}_{g_{j}}&= -w_j X_{\Sigma_g} \alpha_g +  {x_{\mu}}_{g_{j}}.
    \end{align}
\end{subequations}

The updated GMM parameters are obtained by using expectation maximization on these noise estimates. 
% These GMM parameters, along with noisy measurements, become the inputs to the next iteration of the parameter estimation step explained in \eqref{Gen_EIV_Solution1}-\eqref{NM_Updation_1st_deri}.
Subsequently, the obtained GMM parameters along with the noisy measurements are set as inputs to the parameter estimation step explained in \eqref{Gen_EIV_Solution1}-\eqref{NM_Updation_1st_deri}.
% that can approximate $y_{e_{g_S}}$ and $X_{e_{g_S}}$
% After finding the noise expressions, expectation maximization is done to determine the new GMM parameters 
% % (means, standard deviations, and weights)
% % parameters: $\mu_g$, $\sigma_g$, and weights ($\Lambda_g$), 
% that can approximate $y_{e_{g_S}}$ and $X_{e_{g_S}}$.
The noise estimation (using \eqref{EIV-noise-estimates}) and the parameter estimation (using \eqref{Gen_EIV_Solution1}-\eqref{NM_Updation_1st_deri}) are carried out iteratively until an optimal solution is obtained for a particular value of $m$.
% to improve the estimate and arrive at the optimal solution for a particular $m$.
This procedure is then repeated for all $m$ in the range $[1, m_{\text{max}}]$. Finally, the Bayesian information criterion is utilized to select the optimal $m$ ($m^*$), and the corresponding parameter estimate is chosen as the output of EGLE \cite{varghese2022transmission}.

This completes the description of the four state-of-the-art linear estimation methods that are equipped to handle non-Gaussian noises in EIV problems. The subsequent sections of this paper delve into a detailed comparison of these methods across diverse scenarios for the LPE problem. 
We start by providing a brief discussion of the significance of LPE in the power system and its underlying mathematical basis.

\section{Need for Periodic LPE using PMUs}
\label{sec:lpe}

% The values of the line parameters are important for a variety of power system operation and planning studies at the transmission level. 
% For example, they serve as inputs to applications such as dynamic line rating computation, optimal power flow calculation, state estimation, and fault location identification.

% , among others.
% The line parameter values serve as input to applications such as dynamic line rating estimation, optimal power flow calculation, and fault location identification, among others. 
Typically, power utilities have a set of line parameter values for each line in their database. However, the line parameters vary over time due to 
% diverse factors such as 
temperature, humidity, and aging \cite{shi2011transmission1, du2012line1}. Hence, it is important to conduct LPE periodically to update the values in the utility database, as it will then benefit \textit{all} downstream applications that rely on these values. 
% Conducting the line parameter estimation to update the line parameter values in the utility database is termed line parameter estimation (LPE) in this context.

 %The transmission line parameter estimation (TLPE) is selected as the case study for the comparisons. Transmission line parameter estimation concerns estimating the line parameters - resistance, reactance, and shunt susceptance - of a medium-length transmission line model. 

% The variation could be due to environmental features such as temperature or humidity, or it could also be due to other system conditions. Since the knowledge of the line parameters is extremely important for efficient and secure operation of transmission systems, the line parameters need to be estimated from measurements. PMU measurements-based line parameter estimation is the best solution in practice these days \cite{xue2019linear, wehenkel2020parameter, gupta2021compound}. 

\begin{figure}
    \centering
    \includegraphics[width=0.32\textwidth]{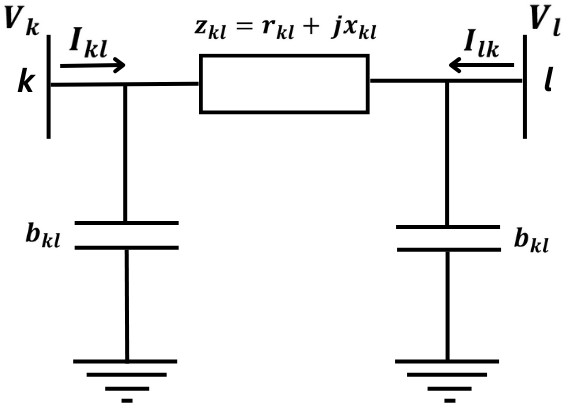}
    \vspace{-1mm}
    \caption{$\pi$-Model for performing LPE}
    \label{fig:pi-model-TL}
\end{figure}

The use of PMU measurements for LPE is extremely popular because of three reasons. 
First, PMUs directly produce voltage and current phasors (magnitude and angle). Hence, they can be used for LPE without any apriori estimation; they also make the LPE problem linear. 
% (as mentioned in the footnote of Section \ref{Intro}). 
Second, the output rate of PMUs is much higher than the speed with which line parameters change. Therefore, PMU-based LPE is
% can be treated as 
a ``static" estimation problem that can be solved off-line;
% with pre-cleaned data; 
judiciously choosing PMU measurement samples also makes the LPE problem linearly independent (rank-sufficient).
Third, as PMU measurements are time-stamped, PMU-based LPE is free from polling and time-skew errors.
The LPE problem is mathematically formulated for a medium-length transmission line whose $\pi$-model
% can be mathematically modeled as follows. The medium length transmission line under consideration can be modeled using a $\pi$  section model 
is shown in Fig. \ref{fig:pi-model-TL}.
The ``from end" and ``to end" of the line are denoted by $k$ and $l$, respectively. The objective of the LPE problem is to update the line parameter values in the utility database using noisy voltage ($\tilde{V}$) and current ($\tilde{I}$) measurements available from PMUs placed at both ends of the line. The series resistance ($r_{kl} \in \mathbb{R}$), series reactance ($x_{kl} \in \mathbb{R}$), and shunt susceptance  ($b_{kl} \in \mathbb{R}$) are the line parameters that must be estimated. Applying Kirchhoff's laws at both ends of the line, the following relationship 
between true voltage phasors, true current phasors, and line parameters 
is obtained,

\vspace{-4mm}

% as shown below
\begin{equation}
\begin{split}
I_{kl_{true}} &= b_k{}_lV_{k_{true}} + (V_{k_{true}} - V_{l_{true}})/z_k{}_l \\
I_{lk_{true}} &= b_k{}_lV_{k_{true}} - (V_{k_{true}} - V_{l_{true}})/z_k{}_l  
\end{split}
\label{kcl}
\end{equation}
where,
% Note that in \eqref{kcl}, 
$z_k{}_l = r_{kl} + jx_k{}_l$. 
% and $x_{kl}$, and can be expressed as
% \begin{equation}
% z_k{}_l = r_k{}_l + jx_k{}_l.
% \end{equation}
% The series admittance, $y_{kl}$, is the inverse of $z_{kl}$. 
By representing the complex currents, voltages, and line parameters of \eqref{kcl} \textcolor{black}{in their Cartesian form}, replacing the true values by their noisy counterparts, and vertically stacking the resulting expressions at different time instances, we obtain the linear regression form shown below,
% \vspace{-1mm}
\begin{equation}
\label{TLPE_noisy}
\begin{bmatrix}
\tilde{I}_{k_r}(1)\\ \tilde{I}_{k_i}(1)\\ \tilde{I}_{l_r}(1)\\ \tilde{I}_{l_i}(1)\\. \\. \\.\\\tilde{I}_{k_r}(t)\\ \tilde{I}_{k_i}(t)\\ \tilde{I}_{l_r}(t)\\ \tilde{I}_{l_i}(t)
\end{bmatrix}\approx
\begin{bmatrix}
\tilde{V}_{k_r}(1) &\tilde{V}_{k_i}(1) &\tilde{V}_{l_r}(1) &\tilde{V}_{l_i}(1)\\
\tilde{V}_{k_i}(1) &$-$ \tilde{V}_{k_r}(1) &\tilde{V}_{l_i}(1) & $-$ \tilde{V}_{l_r}(1)\\
\tilde{V}_{l_r}(1) &\tilde{V}_{l_i}(1) &\tilde{V}_{k_r}(1) &\tilde{V}_{k_i}(1)\\
\tilde{V}_{l_i}(1) &$-$ \tilde{V}_{l_r}(1) &\tilde{V}_{k_i}(1) & $-$ \tilde{V}_{k_r}(1)\\
. &. &. &.\\
. &. &. &.\\
. &. &. &.\\
\tilde{V}_{k_r}(t) &\tilde{V}_{k_i}(t) &\tilde{V}_{l_r}(t) &\tilde{V}_{l_i}(t)\\
\tilde{V}_{k_i}(t) &$-$ \tilde{V}_{k_r}(t) &\tilde{V}_{l_i}(t) & $-$ \tilde{V}_{l_r}(t)\\
\tilde{V}_{l_r}(t) &\tilde{V}_{l_i}(t) &\tilde{V}_{k_r}(t) &\tilde{V}_{k_i}(t)\\
\tilde{V}_{l_i}(t) &$-$ \tilde{V}_{l_r}(t) &\tilde{V}_{k_i}(t) & $-$ \tilde{V}_{k_r}(t)\\
\end{bmatrix}
\begin{bmatrix} 
Y_1\\ Y_2\\ Y_3\\ Y_4
\end{bmatrix}
\end{equation}
where, \begin{math} Y_1 = y_{kl_r}\end{math}, \begin{math} Y_2 = -(b_{kl}+y_{kl_i})\end{math}, \begin{math} Y_3 = -y_{kl_r}\end{math}, \begin{math} Y_4 = y_{kl_i}\end{math}, and \begin{math}y_{kl}\end{math} is the inverse of \begin{math}z_{kl}\end{math}. 
\textcolor{black}{Note that \eqref{TLPE_noisy} is equivalent to \eqref{Noisy_linear_system} in the sense that the dependent variables are the current measurements and the independent variables are the voltage measurements, both of which are obtained from PMUs.
% for this linear regression problem. 
% Finally, \eqref{TLPE_noisy} is made up of noisy voltage and current measurements, whose noise characteristics are represented by some generic non-Gaussian distribution.
Furthermore, since $Y_1 + Y_3 = 0$, \eqref{TLPE_noisy} is also compatible with \eqref{EIV_noisy_linear_system_indep_variable}. 
Lastly, since line parameters lie within $\pm 30\%$ of their database values \cite{kusic2004measurement}, the values in the power utility database are good starting conditions for the iterative techniques described in Section \ref{sec:SOTA}.}
% Lastly, line parameters lie within $\pm 30\%$ of their database values \cite{kusic2004measurement}. This implies that the values in the power utility database are good starting conditions for the iterative techniques described in Section \ref{sec:SOTA}.

Once \eqref{TLPE_noisy} is solved using an appropriate technique, the line parameters can be recovered using the following equation: 
\vspace{-1mm}
\begin{equation}
\begin{split}
r_{kl} &= \frac{2(Y_1-Y_3)}{(Y_1-Y_3)^2+(2Y_4)^2}\\
x_{kl} &= \frac{-4Y_4}{(Y_1-Y_3)^2+(2Y_4)^2}\\
b_{kl} &= -(Y_2+Y_4).
\label{rxb}
\end{split}
\end{equation}

This concludes the mathematical basis of the LPE problem. 
In the next section, we evaluate how the four methods described in Section \ref{sec:SOTA} compare against each other for this problem under diverse measurement noise conditions.

\section{Results} 
\label{sec:perf}

% For evaluating the performance of the LPE using various techniques, the IEEE 118-bus system is selected as the test system. Typically, all the transmission lines in the highest voltage network in a power system are covered with PMU at both ends. Hence, we limit our LPE study to only this voltage level of the power system known as an extra high voltage (EHV) network. 
The IEEE 118-bus system is used to evaluate
% selected as the test system for evaluating 
the performance of MTEE, MTC, CMTC, and EGLE in solving the LPE problem.
As power utilities typically place PMUs on the highest voltage (HV) buses first \cite{varghese2023timesynchronized}, it is assumed that the HV lines of this system have PMUs at both ends.
There are ten such lines in the 118-bus system, and so PMU-based LPE is done for them. 
%The EHV network of the IEEE 118 bus system is shown in Fig. \ref{fig:EHV-118bus}.
%\begin{figure}[H]
%    \centering
%\includegraphics[width=0.45\textwidth]{IEEE_118_bus_test_system_HV_marked.png}
%    \caption{Extra high voltage network of IEEE 118 bus system}
%    \label{fig:EHV-118bus}
%\end{figure}
The data for this study is generated in MATPOWER by simulating the morning load pickup in which the load (and generation) rises steadily \cite{gao2012dynamic}.
%\cite{zimmerman2010matpower}.
% software \cite{zimmerman2010matpower}. 
% The data for this study is generated using MATPOWER. %\cite{zimmerman2010matpower}.
% % software \cite{zimmerman2010matpower}. 
% The morning load pickup is simulated in which the load (and generation) rises steadily.
%\cite{gao2012dynamic}. 
The power flow solutions at different loading conditions provide the true 
% values of the 
voltage and current phasors \textcolor{black}{in p.u}.
The noisy data is created by adding appropriate noise to the true values.
% A morning load pickup pattern is used to create a set of realistic loading conditions. Subsequently, power flows at each of these loading conditions provide us the true values of the voltage and current measurements at various time instants. The noisy voltage and current measurements are created by adding corresponding realistic (non-Gaussian) noise to the corresponding true values. 
Finally, the noisy voltage and current measurements
along with an initial guess of the line parameters
are set as inputs to the four techniques.
% for LPE.

For analyzing the performance of the different estimation techniques, absolute relative error (ARE) is chosen as the performance metric. It is defined as
\vspace{-1mm}
\begin{equation}
w_{\mathrm{ARE}} = \left|\frac{\hat{w} - w_{true}}{w_{true}} \right|
\end{equation}
where, $\hat{w}$ denotes the estimated value of the parameter vector.

\subsection{Evaluation in presence of GMM noise}
\begin{figure}
    \centering
    \includegraphics[width=0.42\textwidth]{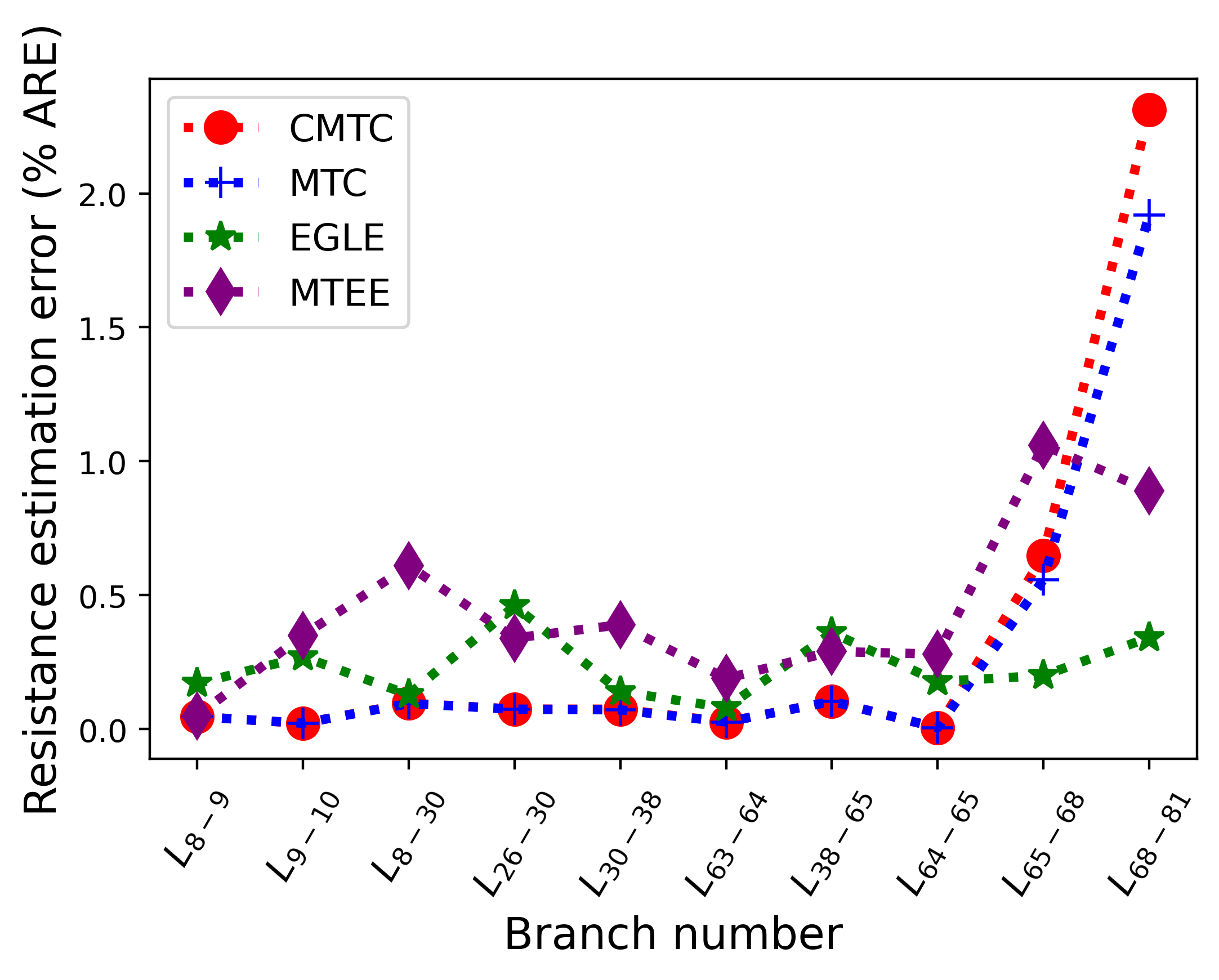}
    \vspace{-5mm}
    \caption{Resistance ARE comparison}
    \label{fig:r_ARE_GMM}
\end{figure}

\begin{figure}
    \centering
    \includegraphics[width=0.42\textwidth]{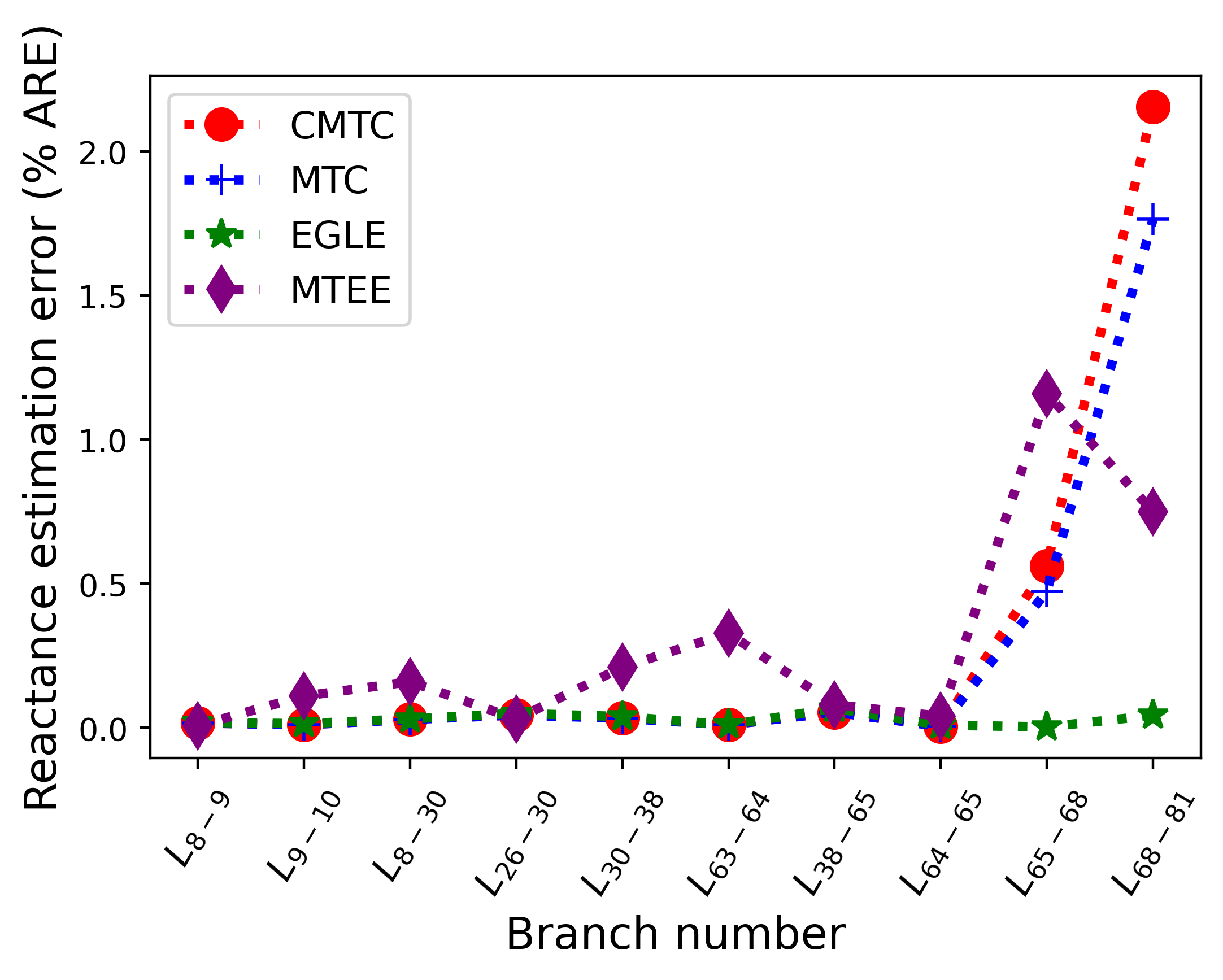}
    \vspace{-3mm}
    \caption{Reactance ARE comparison}
    \label{fig:x_ARE_GMM}
\end{figure}

\begin{figure}
    \centering
    \includegraphics[width=0.42\textwidth]{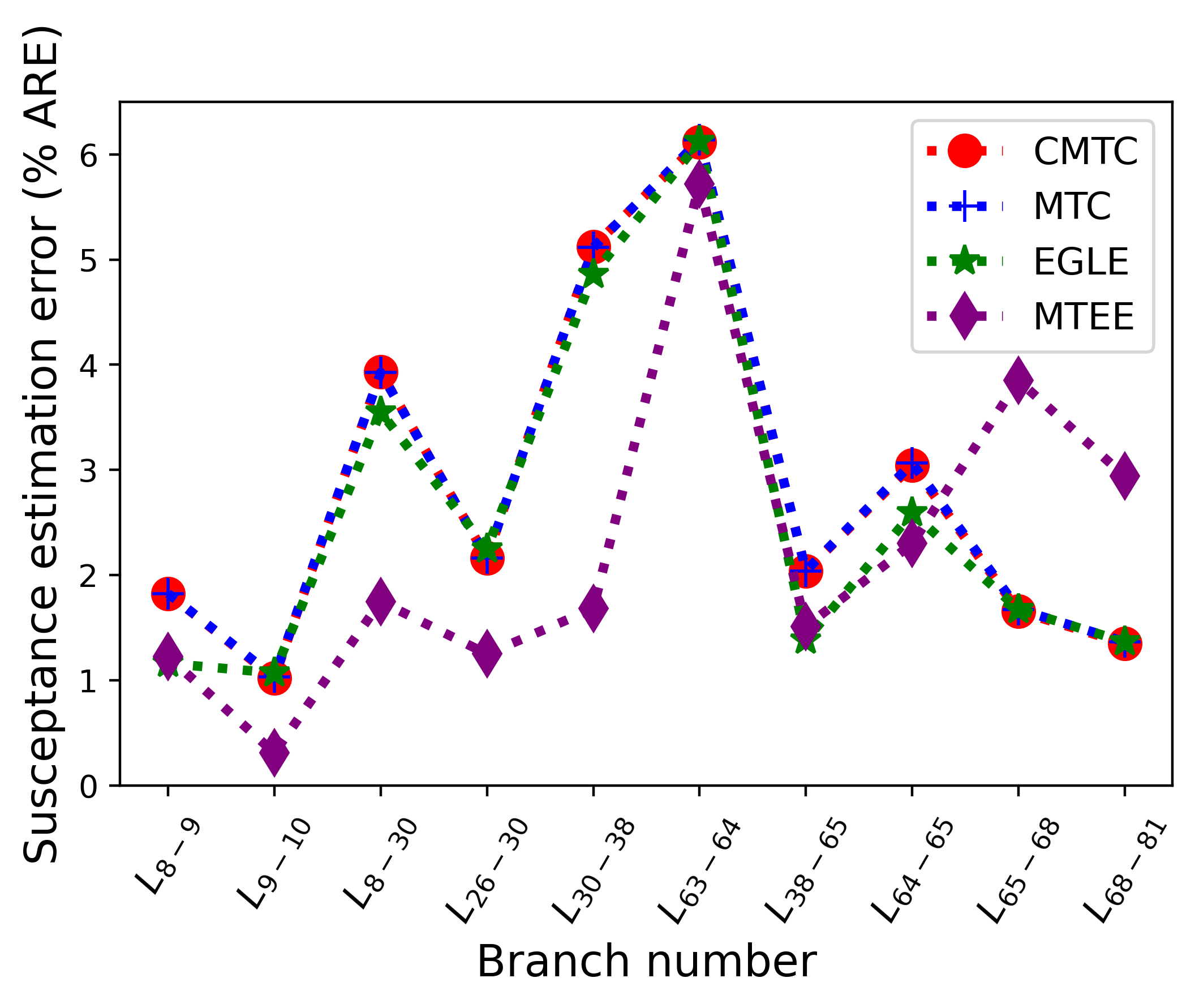}
    \vspace{-3mm}
    \caption{Susceptance ARE comparison}
    \label{fig:b_ARE_GMM}
\end{figure}

For the first study, a non-Gaussian noise in the form of a two-component GMM is used to create the noisy data. The characteristics of the GMM (\textcolor{black}{in p.u.})
% of the GMM noise 
are as follows: $\mu$ = [0, 0.01], $\sigma$ = [0.002, 0.002], $\Lambda$ = [0.3, 0.7], where $\mu$, $\sigma$, and $\Lambda$ represents the mean, standard deviation, and weight vectors of the GMM, respectively.
The estimation error in terms of \%ARE for the four methods is compared in Figs. \ref{fig:r_ARE_GMM}-\ref{fig:b_ARE_GMM}.

\begin{figure}
    \centering
    \includegraphics[width=0.42\textwidth]{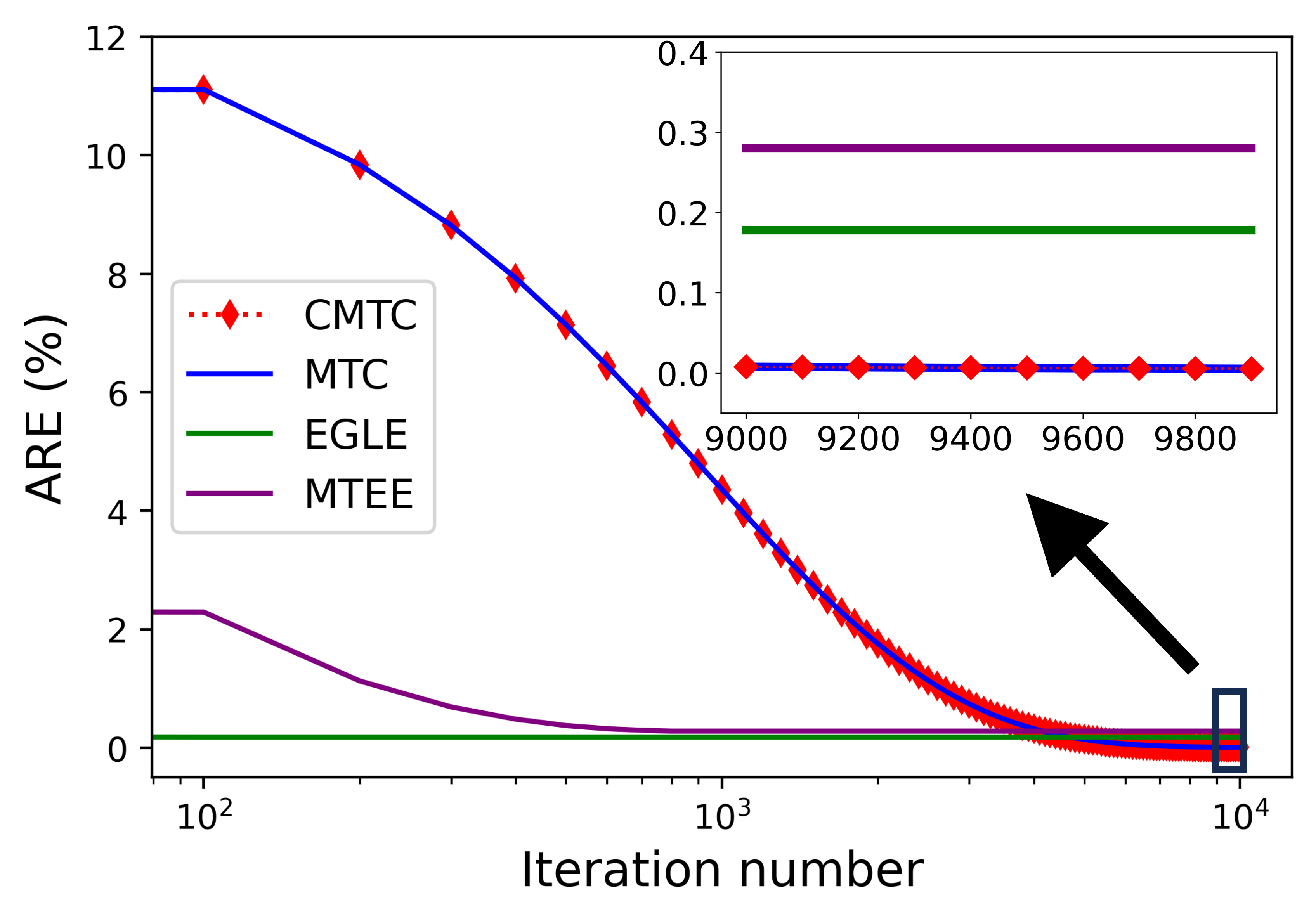}
    \vspace{-3mm}
    \caption{ARE variation with iterations for $L_{64-65}$}
    \label{fig:ARE_with_iterations}
\end{figure}

Figs. \ref{fig:r_ARE_GMM} and \ref{fig:x_ARE_GMM} compare the estimation error for resistance and reactance estimates, respectively.
It can be observed that all four methods provide very good estimates under non-Gaussian measurement noise for most branches. 
% The reactance estimates displayed in Fig. \ref{fig:x_ARE_GMM} also demonstrate a similar behavior. 
One possible reason for the last two branches ($L_{65-68}$ and $L_{68-81}$) having a higher error could be the relatively poorer conditioning of the EIV problem form of these two lines in comparison to the other lines.
% The reason for the last two branches ($L_{65-68}$ and $L_{68-81}$) having a higher error 
% % for both resistance and reactance estimates 
% is due to their EIV problem form being more ill-conditioned than the other lines.
% the ill-conditioning of the EIV problem for those two lines.
% except for the two branches with high condition number. The reasoning explained for resistance estimates holds true here as well. 
The estimation accuracy of the shunt susceptance is compared in Fig. \ref{fig:b_ARE_GMM}. 
It can be observed from the figures that MTEE generally has better accuracy for the susceptance estimates, whereas MTC, CMTC, and EGLE have lower errors for the resistance and reactance estimates.
% It can be observed that the MTEE has the highest accuracy for the shunt susceptance estimation followed by EGLE. This is in contrast to the slightly better estimation accuracy of MTC and CMTC with respect to the MTEE for the resistance and reactance estimates.
The relatively high error values for the susceptance estimates
% (in comparison to the resistance and reactance estimates) 
can be explained as follows: even though $Y_2$ and $Y_4$ are individually estimated with high accuracy, when they are combined to calculate susceptance using \eqref{rxb}, the errors 
% in the estimates 
could get added up to result in a high net \%ARE for the susceptance estimates.
This problem can be tackled by creating alternate forms of \eqref{TLPE_noisy}, and will be explored in the future. To analyze convergence behavior of the four methods, \%ARE of resistance estimates w.r.t. number of iterations are plotted for $L_{64-65}$ in Fig. \ref{fig:ARE_with_iterations}. EGLE is found to converge within very few iterations ($\approx$20 iterations), MTEE takes around $10^3$ iterations, while MTC and CMTC keep improving until about $10^4$ iterations. Similar convergence behavior was observed for reactance and susceptance estimates as well.
% \textcolor{black} {The susceptance ARE is higher than the reactance and resistance values this can be attributed to the correct estimation of $Y_2$ and $Y_4$ individually, however when combined for the calculation of susceptance as shown in \eqref{rxb}, the additive resultant of these components is higher than the resistance and reactance ARE values.}
 
%EGLE is found to converge within very few iterations ($\approx$30 iterations), MTEE takes around $10^3$ iterations, while MTC and CMTC keep improving until about $10^4$ iterations. Similar convergence behavior was observed for reactance and susceptance estimates as well. 
%However, the time required per iteration for MTEE was \textit{approximately four to five orders of magnitude higher} than the others.
% MTC/CMTC/EGLE.
%This is due to the double summation that must be computed over all the samples (see \eqref{MTEE_update}), implying that MTEE can only be used when there is sufficient time and compute resources.
% \textcolor{black}{Although the time required per iteration for MTEE is five order of magnitude times the MTC suggesting that MTC could be employed for quick estimation.} 
% The performance of the four methods in presence of other types of measurement noise are described next. 

\subsection{Evaluation in presence of other noise distributions}
% Noise characterization of PMU measurements is an active area of research. Some literature claims that Laplacian distribution is a better way to model the PMU noise, and some others show under certain conditions Gaussian distribution is sufficient to model the noise.  
In this subsection, we evaluate performance of the four methods under Laplacian and Gaussian noise conditions. 
The Laplacian noise has a mean of 0 and a scale of 0.005.
The Gaussian noise has zero mean with a standard deviation of 0.005. 
The results obtained for $L_{64-65}$ \textcolor{black}{along with the time taken to do so,} are displayed in Tables \ref{Lapl_noise_ARE_comaprison} and \ref{Gauss_noise_ARE_comaprison}, respectively, in which $\mathrm{TV}$ refers to the \textit{true value}.
All four methods produce reasonably good estimates for the Laplacian noise case; this confirms that they can effectively handle different types of non-Gaussian noises in PMU measurements.
For the Gaussian noise case, the four techniques were also compared with TLS. 
\textcolor{black}{It was observed from the two tables that EGLE had the best performance while also being the quickest among the four techniques.} 
% However MTEE was \textit{approximately four to five orders of magnitude higher} than the others. This is due to the double summation that must be computed over all the samples (see \eqref{MTEE_update}), implying that MTEE can only be used when there is sufficient time and compute resources.
Particularly, MTEE took \textit{approximately four orders of magnitude} longer time than the other methods. This was due to the double summation that must be computed over all the samples (see \eqref{MTEE_update}), implying that MTEE can only be used when there is sufficient time and compute resources.
% It can be observed that all methods have very good accuracy under the Gaussian noise case. This reassures the knowledge that even when the noise reduces to a Gaussian-like scenario, these methods produce reliable estimates. Comparing the four methods the results obtained for EGLE are exceptional. 
% The methods produce reasonably good estimates for the Laplacian noise case as well. This goes on to show that the methods are reliable for any type of noise (including Gaussian noise and variety of non-Gaussian noises) that may occur. 

% The next subsection discusses the peculiar features and characteristics of each of the four algorithms. 
%\subsection{Evaluation in presence of Gaussian Noise}

\subsection{Discussion}
% It was observed that all four methods provided accurate line parameter estimates in general for the non-Gaussian measurement noise. It was also shown that they performed equally well when the measurement noise characteristics were reduced to Gaussian. Consequently, each of the four methods - MTC, CMTC, EGLE, and MTEE - proves suitable for line parameter estimation.
The analyses conducted in the previous subsections confirmed that MTEE, MTC, CMTC, and EGLE can be used to perform LPE in presence of any type of noise in the PMU measurements.
% Subsequent sections elucidate distinct features of these methods, guiding power system operators in selecting the most appropriate one based on specific conditions. 
In this subsection, we identify conditions under which these techniques are expected to give good results.
This will be extremely beneficial to power system operators as it will help them select the appropriate method for their problem.
%Subsequent paragraphs elucidate distinct features of these methods that could help power system operators in selecting the most appropriate one based one for line parameter estimation or even other linear estimation problems.

\vspace{-1em}

\begin{table}[H]
\centering
\caption{Estimates in p.u. under Laplacian noise for $L_{64-65}$}
\vspace{-3mm}
\label{Lapl_noise_ARE_comaprison}
\begin{tabular}{|l|c|c|c|c|c|}
\hline
  & TV & MTC  & CMTC & EGLE  & MTEE \\ \hline
r &  0.00269  & 0.00267 & 0.00267 & 0.00269  & 0.00268 \\ \hline
x &  0.0302  & 0.0300 & 0.0301 & 0.0302 & 0.0301 \\ \hline
b &   0.3800 & 0.3797 & 0.3798 & 0.3797  & 0.3796 \\ \hline 
%\textcolor{black}{Time (s)} &    & 10.07  & 8.42 &  239  & 78000 \\ \hline
%\textcolor{black}{Time (s)} & \multicolumn{2}{l|}{}  & 8.42 &  239  & 78000 \\ \hline
\multicolumn{2}{|c|}{\textcolor{black}{Time (s)}} & \textcolor{black}{4.07} & \textcolor{black}{4.24} &  \textcolor{black}{1.68}  & \textcolor{black}{7.3e+4} \\ \hline
%\multicolumn{2}{\textcolor{black}{Time (s)}}{} & 10.07 & 8.42 &  239  & 78000 \\ \hline

\end{tabular}
\end{table}

\vspace{-2em}

\begin{table}[H]
\centering
\caption{Estimates in p.u. under Gaussian noise for $L_{64-65}$}
\vspace{-3mm}
\label{Gauss_noise_ARE_comaprison}
\begin{tabular}{|l|c|c|c|c|c|c|}
\hline
   & TV  & TLS    & MTC  & CMTC & EGLE  & MTEE  \\ \hline
%r  & 0.00269 & 0.00269    & 0.00271 & 0.00271 & 0.00269 & 0.00270  \\ \hline
r  & 2.69e-3 & 2.69e-3    & 2.71e-3 & 2.71e-3 & 2.69e-3 & 2.70e-3  \\ \hline
x  & 30.2e-3 & 30.2e-3    & 30.4e-3  & 30.4e-3  & 30.2e-3 & 30.2e-3  \\ \hline
%x  & 0.0302 & 0.0302    & 0.0304  & 0.0304  & 0.0302 & 0.0302  \\ \hline
b  & 0.3800 & 0.3801   & 0.3801 & 0.3801 & 0.3800 & 0.3801  \\ \hline
%\textcolor{black}{Time (s)} &  & 88   & 7.6 & 7.7 & 239 & 78000 \\ \hline
\multicolumn{2}{|c|}{\textcolor{black}{Time (s)}} & \textcolor{black}{0.35} & \textcolor{black}{4.14} &  \textcolor{black}{4.18}  & \textcolor{black}{1.68} & \textcolor{black}{7.3e+4} \\ \hline
\end{tabular}
\end{table}

%Next, some features that differentiate these methods are explained that would help the power system operator in choosing one method over another based on the conditions.
%One challenge faced while implementing these estimation techniques is the need to tune the parameters such as learning rate and kernel width. In MTEE, MTC, and CMTC algorithms the choice of learning rate plays a crucial role in achieving both accuracy as well as speed. To optimize outcomes, rigorous tuning of the learning rate is essential when deploying these three methods. Given that line parameter estimation typically starts with a reasonable initial estimate (sourced from the utility database), a practical guideline is to set the learning rate at one-hundredth of the initial guess's cost function for MTEE, MTC, and CMTC. Notably, the EGLE algorithm stands out as it doesn't necessitate parameter tuning across different branches, enabling it to deliver fast and accurate results without any tuning of the parameters.

\textit{Remark 1: Tuning.} One challenge faced while implementing the ITL techniques (MTEE, MTC, and CMTC) is the need to tune the hyperparameters, such as step size and kernel width. 
Specifically, for all three ITL techniques, the step size was found to play a crucial role in achieving both accuracy as well as speed. Hence, it had to be independently tuned for every line.
In this regard, the EGLE algorithm stands out as it does not need to be separately tuned for the different lines.
% In MTEE, MTC, and CMTC algorithms the choice of learning rate plays a crucial role in achieving both accuracy as well as speed. Notably, the EGLE algorithm stands out as it doesn't necessitate parameter tuning across different branches, enabling it to deliver fast and accurate results without any tuning of the parameters.

\textit{Remark 2: Applicability.} EGLE had a lower efficacy when the initial guess was far away from the actual values of the line parameters.
However, given that the variability of the  values is constrained to be within established limits ($\pm 30\%$ \cite{kusic2004measurement}), this is not a major concern for the LPE problem. 
% relied heavily on the proximity of the initial guess of the line parameters to the true/actual values. 
% Given that the variability of the line parameter values is constrained within established limits ($\pm 30\%$ \cite{kusic2004measurement}), this is not a major concern for the LPE problem.
% When
% However, when considering its application to other problems that may 
For applications that do not have such constraints on the parameters, modifications to the EGLE technique may be necessary. 
Conversely, MTEE, MTC, and CMTC exhibited less sensitivity to initial conditions.
Hence, a potential approach could be to first solve the estimation problem using MTC (with zero initial conditions) and then use the MTC results as an input to EGLE.

\textit{Remark 3: Operating guidelines.} 
% Finally, a comparison of the methods based on the underlying principles is carried out. 
In MTEE, the quadratic Renyi's entropy captures the uncertainty 
% or disorder 
in the error distribution, which can lead to better generalization in scenarios where the underlying distribution is complex. 
% or uncertain. 
% On the other hand, 
MTC's use of the correntropy function enhances its robustness against outliers. 
% The correntropy function's diminished penalty on substantial deviations, compared to squared error, renders it less vulnerable to outliers. 
Specifically, the function's reduced sensitivity to substantial deviations compared to squared error plays a pivotal role in enhancing the speed of MTC.
This could not be demonstrated in the identified application because LPE is typically done offline using pre-cleaned (outlier-free) data. 
% (obtained, say, from a PMU-based state estimator).
CMTC retains the merits of MTC, with its additional equality constraint further boosting the accuracy under appropriate scenarios.
% Meanwhile, 
EGLE, which models noise as a GMM followed by parameter estimation, is both anticipated and demonstrated to proficiently manage diverse noise distributions with minimal user effort.

 \section{Conclusion} 
 \label{sec:conc}

Four techniques are investigated in this paper to solve the PMU-based LPE problem. This problem is a linear EIV problem with the possibility that the noises in the dependent and independent variables have non-Gaussian distributions. 
Each method  was found to be competent for this task.
The accuracy of MTEE was a function of rigorous hyperparameter tuning as well as availability of sufficient time and compute resources.
% Each method was found to be competent for the task, with the MTEE method demonstrating a slightly superior estimation accuracy. This heightened accuracy, however, necessitated rigorous hyperparameter tuning as well as sufficient time and compute resources.
MTC/CMTC are commendable alternatives to the (computational burden of the) MTEE technique; however, they have comparatively lower estimation accuracy.
% However, their accuracy was found to be low for the LPE problem.
The EGLE method yields fast and accurate results without the need for any hyperparameter tuning.
% parameter adjustments. 
Its effective performance hinges on intelligent initialization, which requires domain knowledge.

\bibliographystyle{IEEEtran}
%\bibliography{TLPE_PS_Ref}

\bibliography{bibilography1}

\end{document}